\documentclass[pra,aps,floatfix,tightenlines,twocolumn,showpacs]{revtex4}
\usepackage{amsmath}
\usepackage{amssymb}
\usepackage{amsfonts}
\usepackage{epsfig}
\usepackage{subfigure}
\usepackage[dvips]{color}
\usepackage{bm}

\begin{document}

\title{Time-correspondence differential ghost imaging}
\author{Ming-Fei Li}
\author{Yu-Ran Zhang}
\author{Kai-Hong Luo}
\author{Ling-An Wu}
\email{wula@iphy.ac.cn}
\author{Heng Fan}
\email{hfan@iphy.ac.cn} \affiliation{Laboratory of Optical Physics, Institute
of Physics and Beijing National Laboratory for Condensed Matter Physics,
Chinese Academy of Sciences, Beijing 100190, China}

\date{revised proof: Mar 12, 2013}

\begin{abstract}
Experimental data with digital masks and a theoretical analysis are presented for an imaging scheme that
we call time-correspondence differential ghost imaging (TCDGI). It is shown that by conditional averaging
of the information from the reference detector but with the negative signals inverted, the quality of the
reconstructed images is in general superior to all other ghost imaging (GI) methods to date. The advantages of both
differential GI and time-correspondence GI are combined, plus less data manipulation and shorter computation
time are required to obtain equivalent quality images under the same conditions. This TCDGI method offers
a general approach applicable to all GI techniques, especially when objects with continuous gray tones are
involved.
\end{abstract}
\maketitle Since the first ``ghost image'' was observed with entangled photon
pairs generated by spontaneous parametric down-conversion \cite{1st}, ghost
imaging (GI) has become a focus of great attention as well as contention. In
this technique, two spatially correlated beams are used to reconstruct the
object image. The object beam passes through the object and its total intensity
is collected by a ``bucket" detector with no spatial resolution; a reference
beam that does not interact with the object is measured by a pixel array
detector. Various radiation sources may be employed, including quantum optical,
pseudo-thermal \cite{t6}, and true thermal light \cite{t7}, while lensless configurations \cite{t8} and even
systems using a computer generated thermal field with a single bucket detector \cite{Shapiro, t11} have been demonstrated, stirring up a fundamental debate
on whether GI is an intrinsically quantum phenomenon or whether it can be
interpreted by classical optics \cite{t9,debate1,debate2}. This
notwithstanding, GI displays great potential because it allows imaging of the
object in harsh environments, e.g. in a scattering medium
\cite{t2} or turbulent atmosphere \cite{cj,t3,t4}, where standard imaging
methods fail. Computational GI may also be used in optical encryption \cite{Clemente-OL}, and in ghost holography, where both intensity and phase information may be retrieved as well as in ghost holography, where
both intensity and phase information may be retrieved \cite{Clemente-SPGH}. The disadvantage is that very long measurement times are needed, while the visibility and signal-to-noise ratio (SNR) are low,
especially with thermal light, which are serious drawbacks for practical
applications. Although compressed sensing \cite{t10} can be used to reduce
the number of measurements required for image reconstruction, or equivalently,
greatly improve the image quality for the same number of exposures in GI, the
corresponding data processing time is also greatly increased.

Recently, another method called differential ghost imaging (DGI) \cite{DGI} was demonstrated
which can dramatically enhance the SNR of conventional GI, but again with a
huge amount of measurement data and more complex computation. Recently, Luo and co-workers
\emph{et al.} \cite{aip,khl} reported a technique that they called
correspondence imaging (CI), in which a positive or negative image is retrieved by conditional
averaging of the reference signals; that is, only those reference data that
correspond to positive or negative intensity fluctuations of the bucket signal are
selected for simple averaging, without the need to multiply by the bucket
detector intensity itself.
Compared with conventional GI for the same number of exposures, the processing time is greatly reduced, since, computationally, addition is faster than multiplication, whilst fewer frames are required to reconstruct the images. Moreover,
the SNR of the negative image of CI is always better than that in conventional
GI, but for the positive images it depends on the partition weighting
\cite{khl}.

In this paper we present another approach that we call
time-correspondence differential ghost imaging (TCDGI), in
which the advantages of DGI and CI are combined. Classical
explanations of the phenomena are presented and we show
that the image reconstructed by TCDGI can be as good as or
even better than that of DGI, but with less data manipulation
required and shorter computation times. This feature is a
definite advantage and represents a step forward towards real
practical applications.

\begin{figure}[!hbt]
\includegraphics[width=0.40\textwidth]{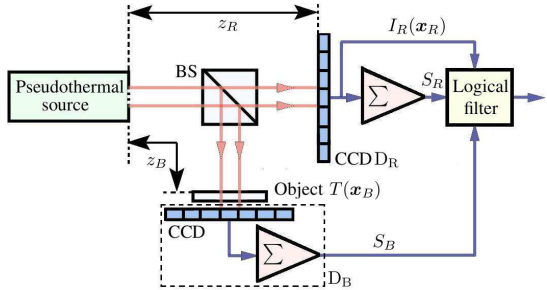}
\caption{(Color online) Schematic diagram of the experimental setup.
\label{setup}}
\end{figure}

The experimental setup, shown in Fig.\ \ref{setup}, is a lensless GI system. A
linearly polarized $632.8$-$\rm{nm}$ He-Ne laser beam is projected onto a
ground-glass disk rotating at 3 rad/min to produce a field of randomly varying
speckles, which have an average diameter of $\delta_{0}\simeq20$ $\rm{\mu m}$.
This pseudo-thermal light is divided by a 50:50 beamsplitter (BS) into two
spatially correlated object and reference beams; the former emerges from the
object with an intensity distribution of $I_{B}(\bm{x}_{B})$ to be collected by
the bucket detector $\rm{D_{B}}$, while the latter arrives at the reference
detector $\rm{D_{R}}$ with a distribution of $I_{R}(\bm{x}_{R})$, where
$\bm{x}$ is the transverse spatial coordinate and the suffixes $\rm B$ and $\rm
R$ represent bucket and reference detectors, respectively. Both beams are collected by identical charge-coupled device (CCD) cameras (Imaging Source DMK
31BU03), of pixel size $4.65$ $\rm{\mu m}$, which are synchronously triggered by a pulse generator. The area of the beams at the object
and reference detector planes, which are at the same distance $z_{B}=z_{R}=215$
$\rm{mm}$ from the source, is $A_{beam}\simeq0.55$ $\rm{mm^{2}}$, and contains
$N_{speckle}=A_{beam}/A_{coh}\simeq1400$ speckles; where we have taken the
coherence area to be $A_{coh}\sim\delta_{0}^{2}$.

\begin{figure}[!hbt]
\centering \subfigure[]{
\includegraphics[width=0.1\textwidth]{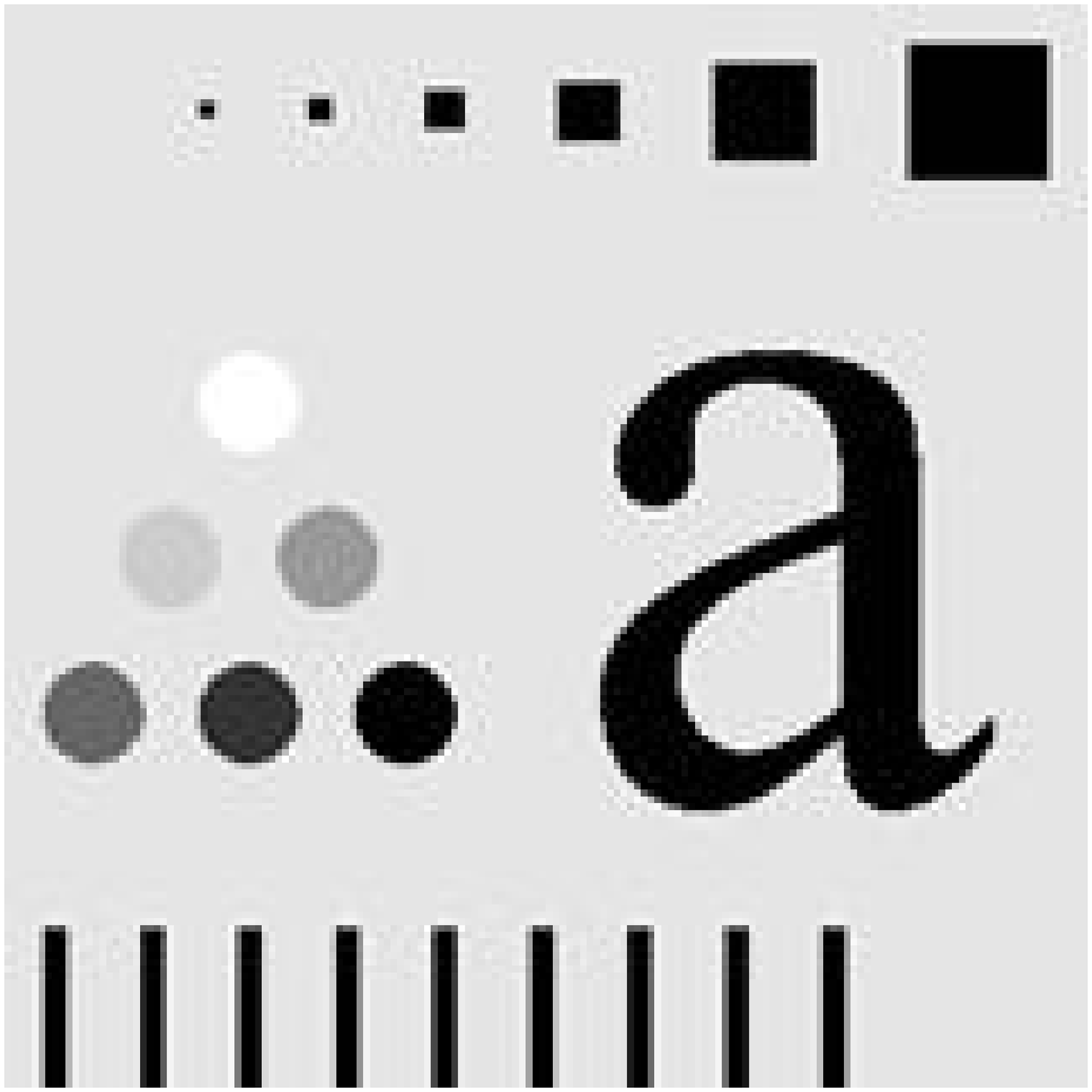}\label{f2a}}
\subfigure[]{
\includegraphics[width=0.1\textwidth]{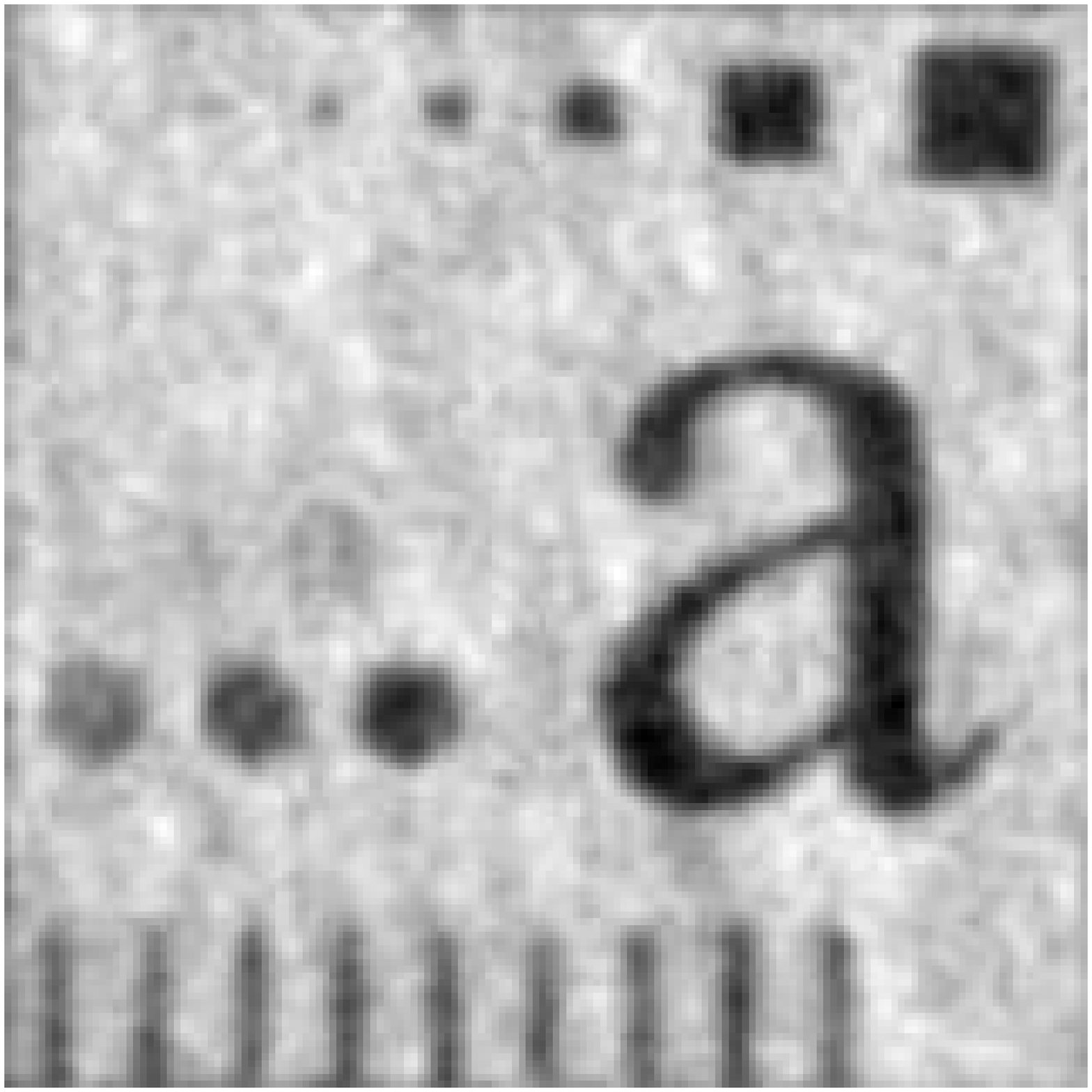}\label{f2b}}
\subfigure[]{
\includegraphics[width=0.1\textwidth]{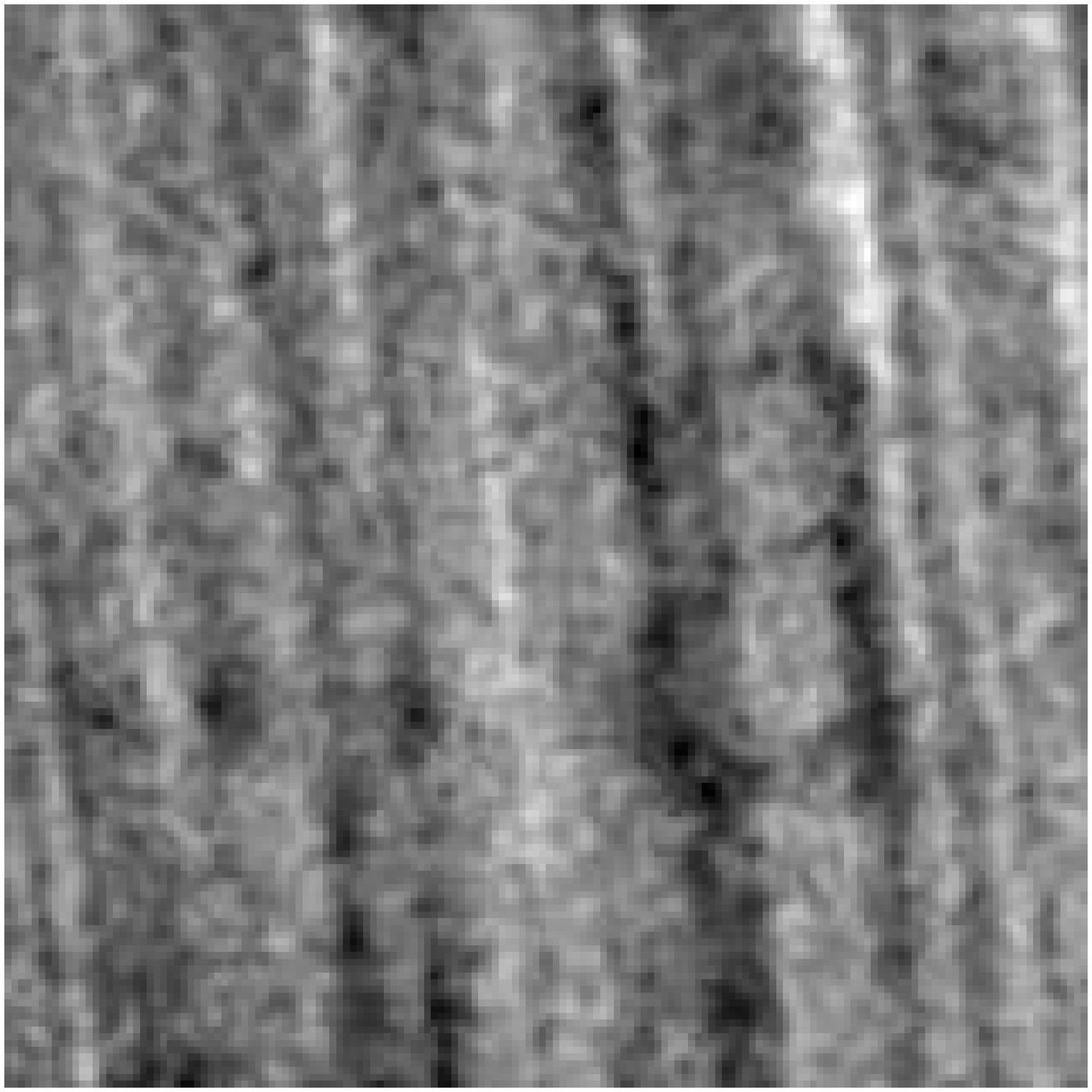}\label{f2c}}
\subfigure[]{
\includegraphics[width=0.1\textwidth]{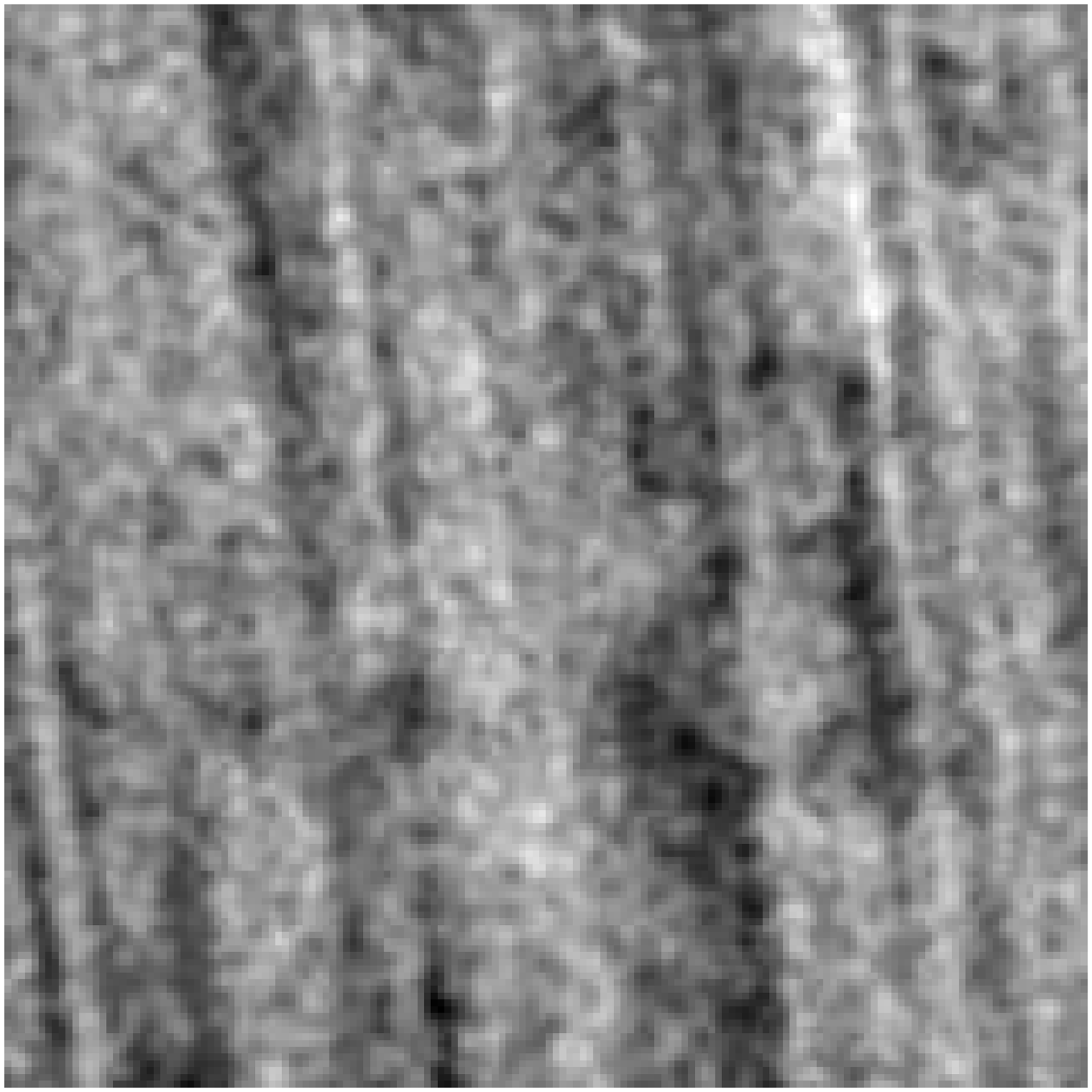}\label{f2d}}
\includegraphics[width=0.02759\textwidth]{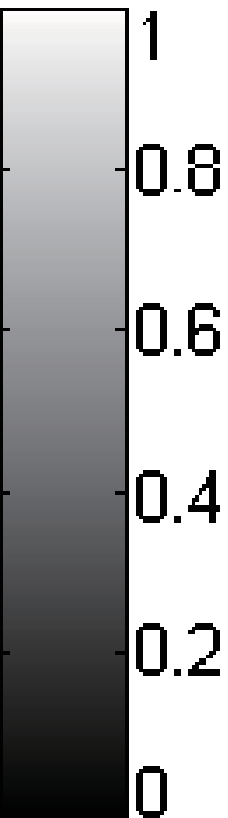}\\
\subfigure[]{
\includegraphics[width=0.1\textwidth]{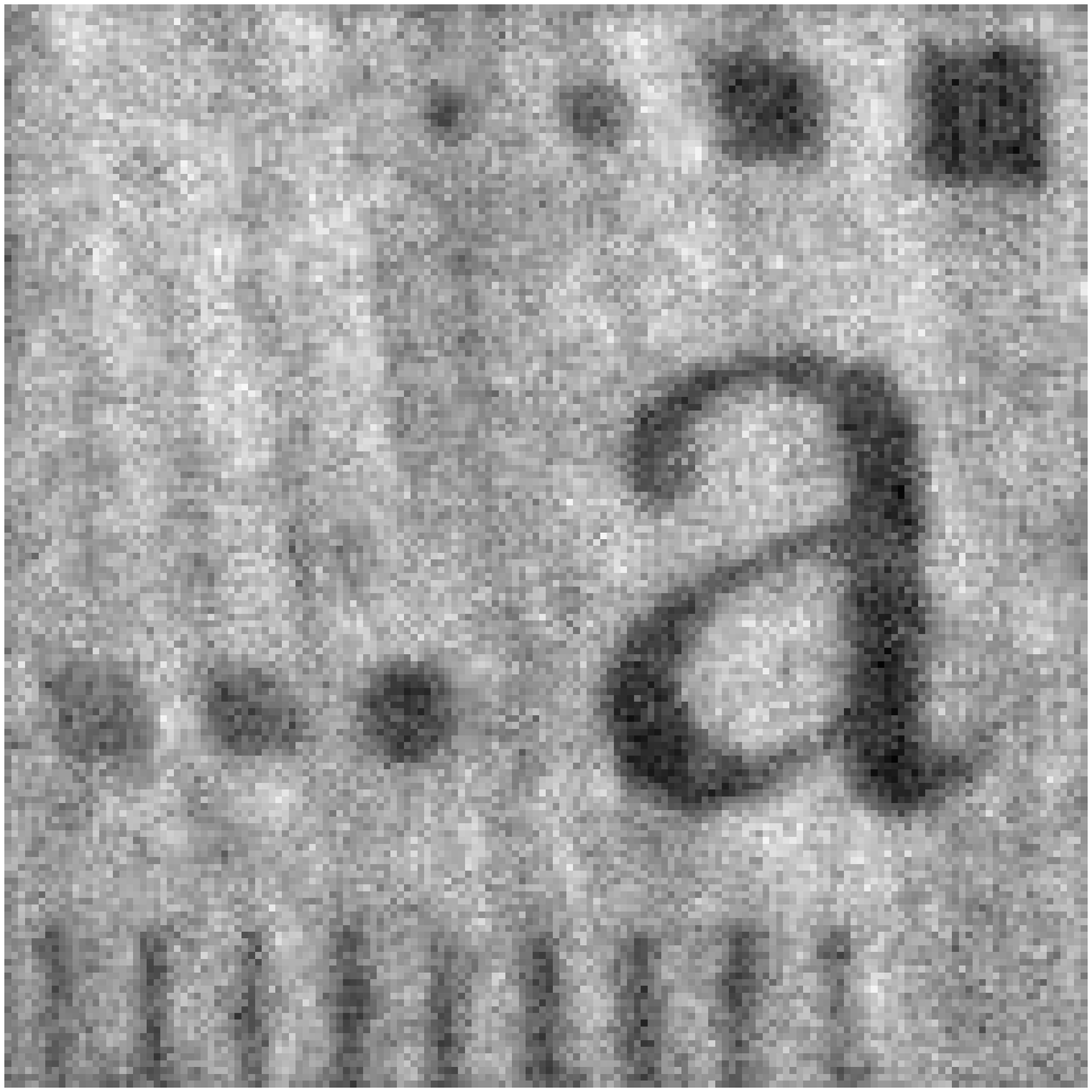}\label{f2e}}
\subfigure[]{
\includegraphics[width=0.1\textwidth]{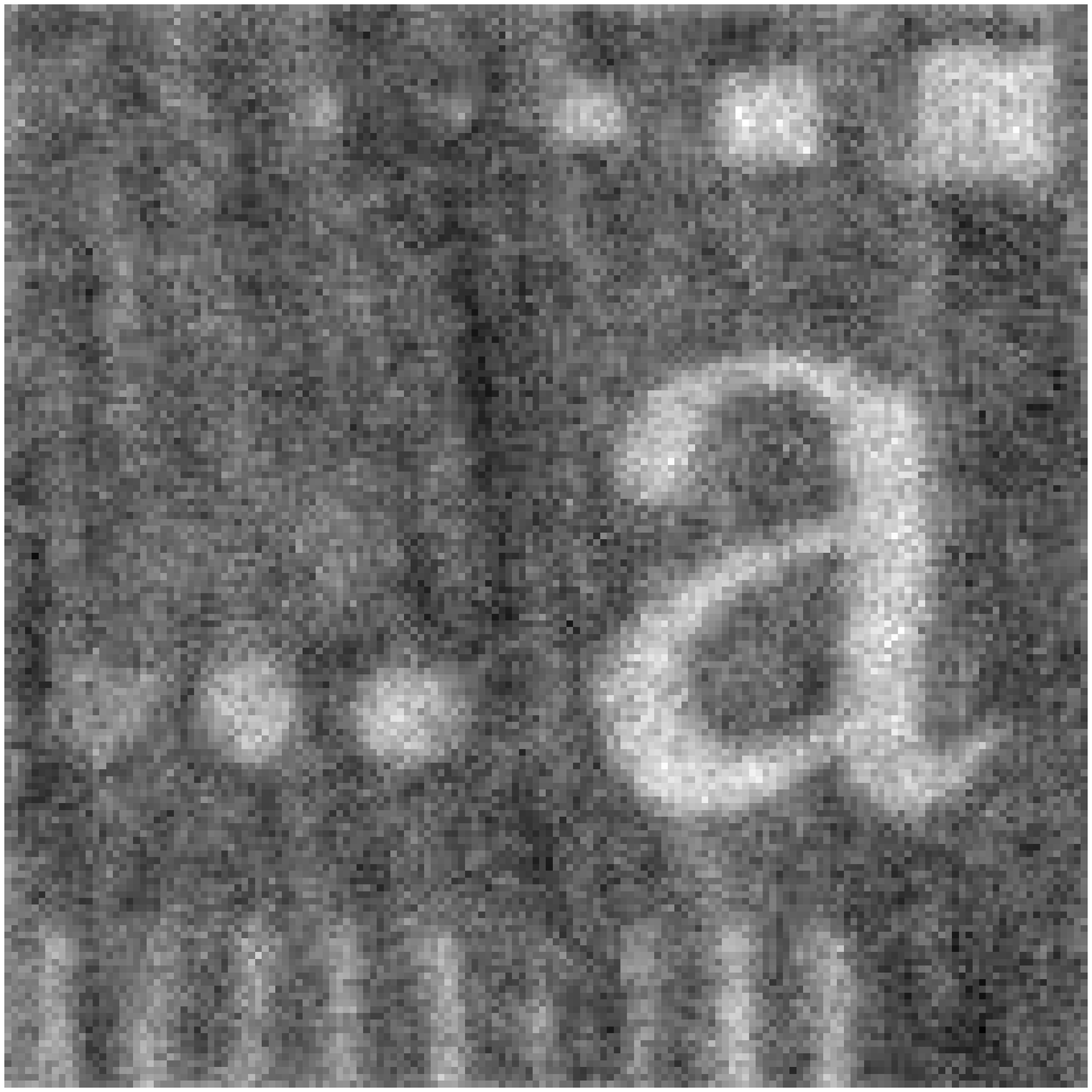}\label{f2f}}
\subfigure[]{
\includegraphics[width=0.1\textwidth]{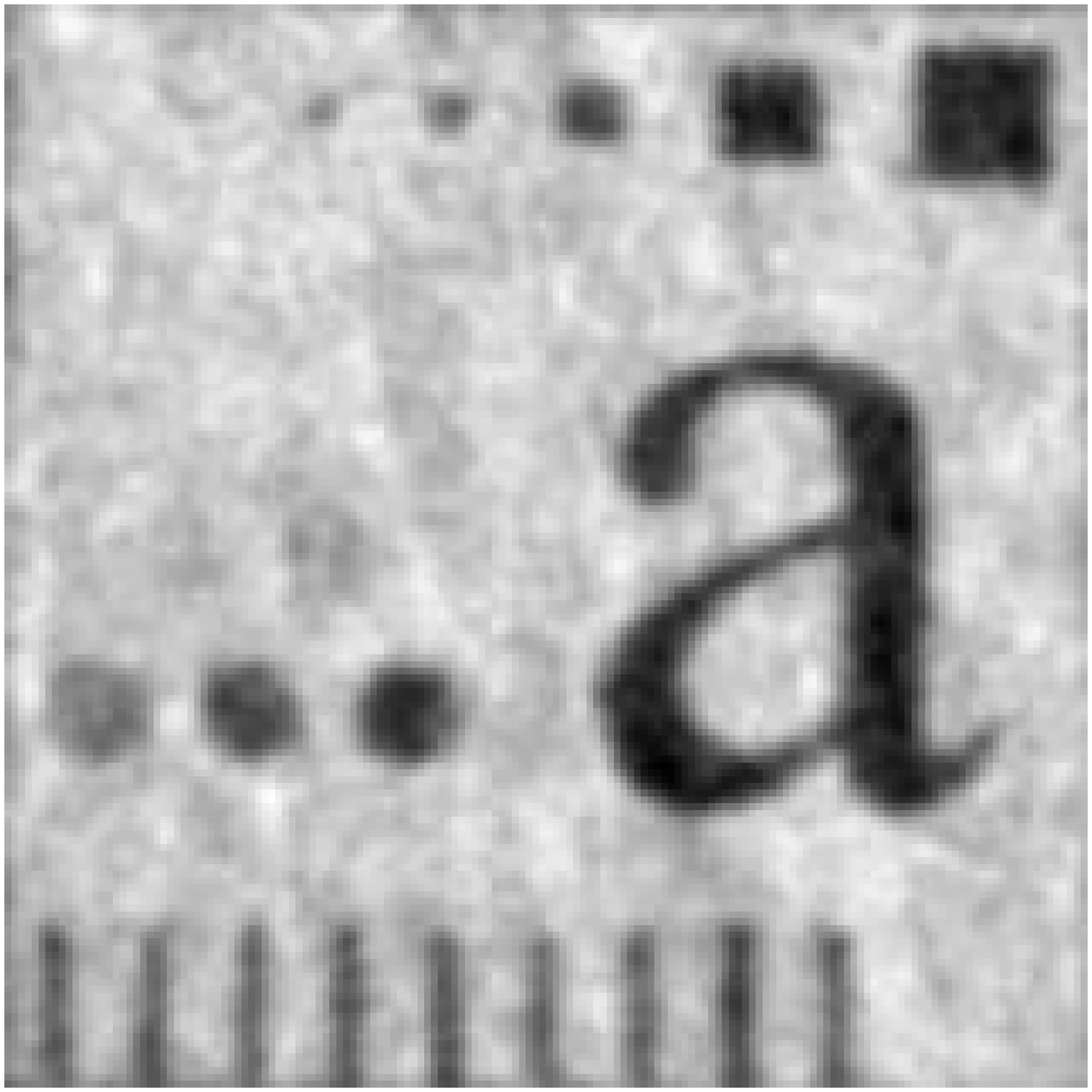}\label{f2g}}
\subfigure[]{
\includegraphics[width=0.1\textwidth]{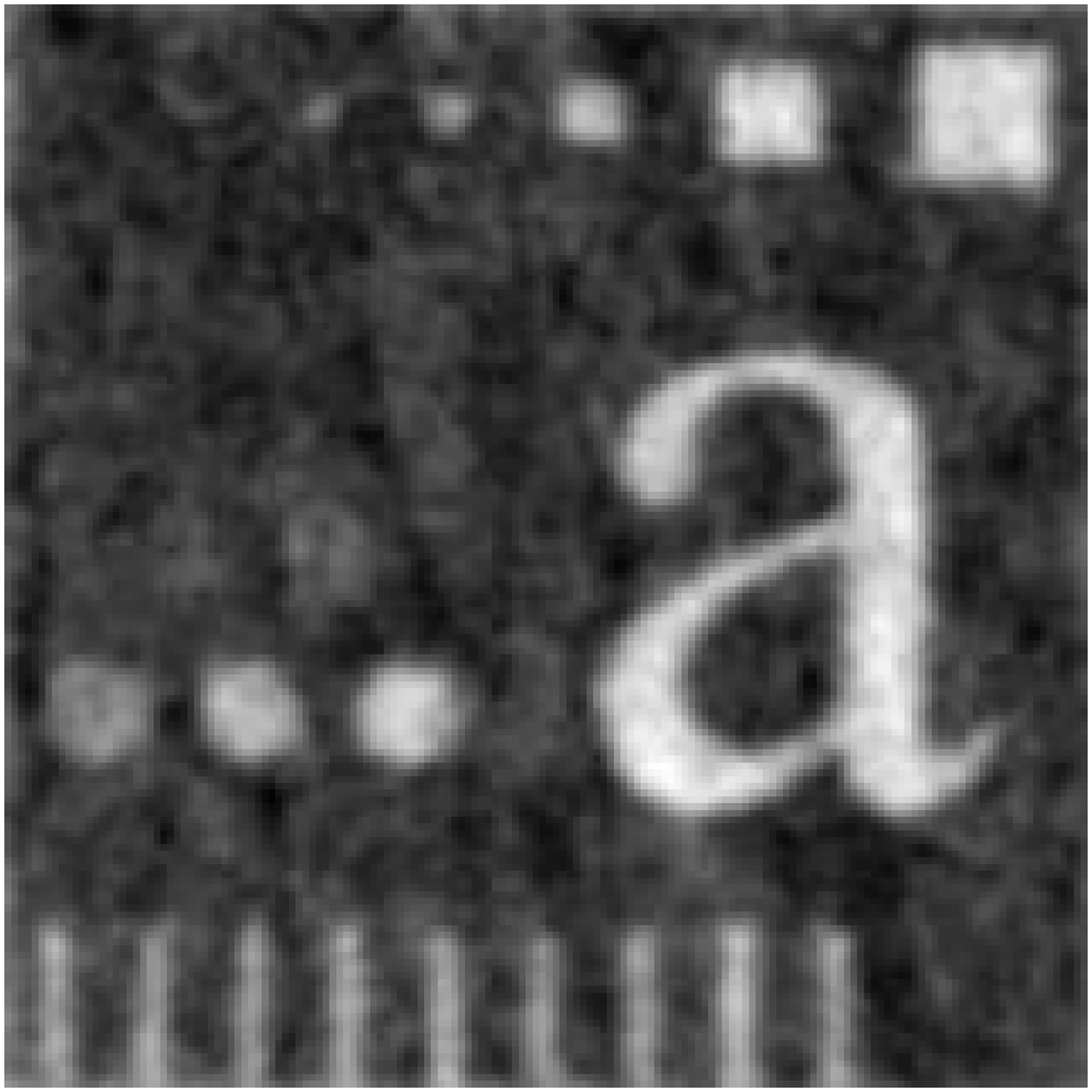}\label{f2h}}
\includegraphics[width=0.02759\textwidth]{figgray2.eps}\\
\subfigure[]{
\includegraphics[width=0.1\textwidth]{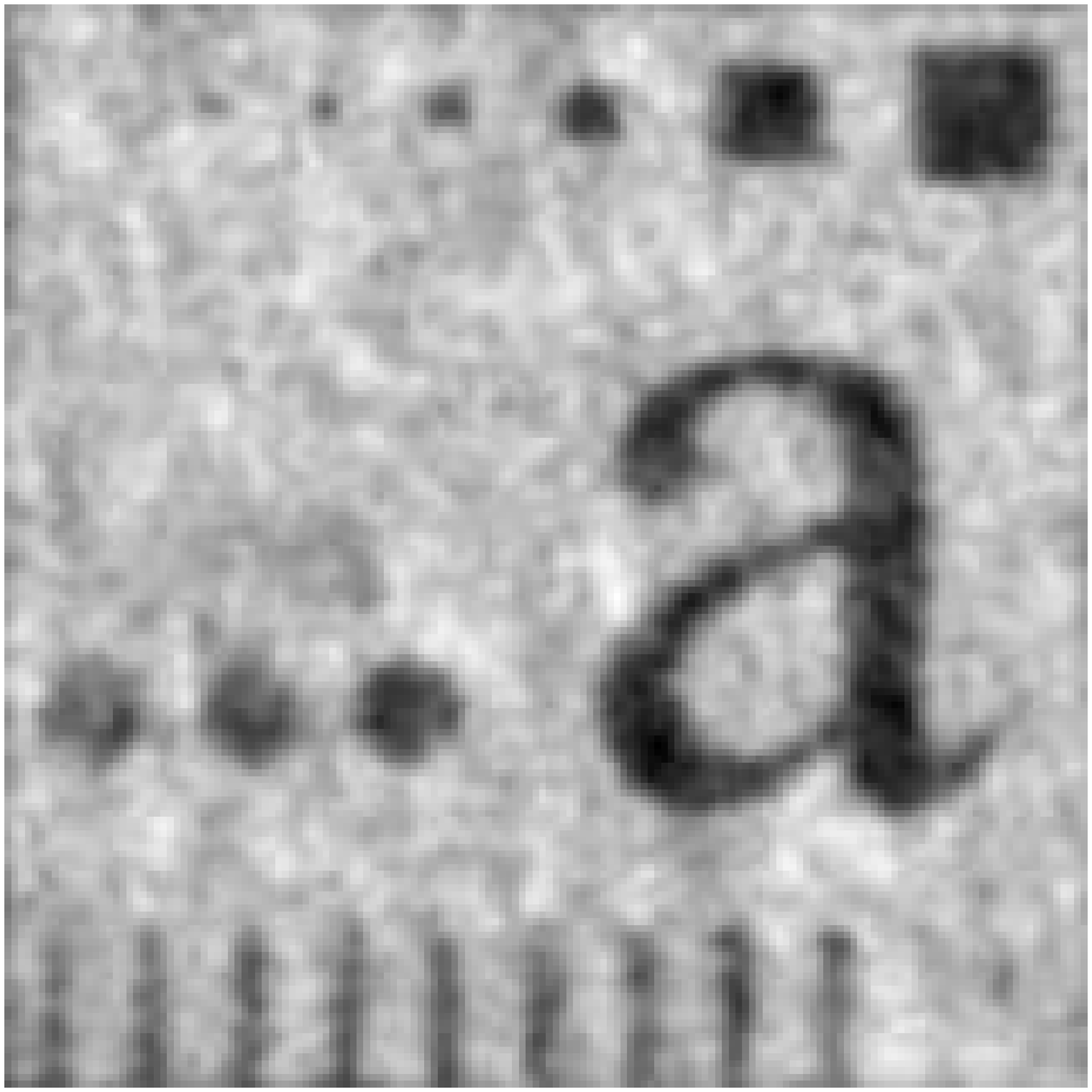}\label{f2i}}
\subfigure[]{
\includegraphics[width=0.1\textwidth]{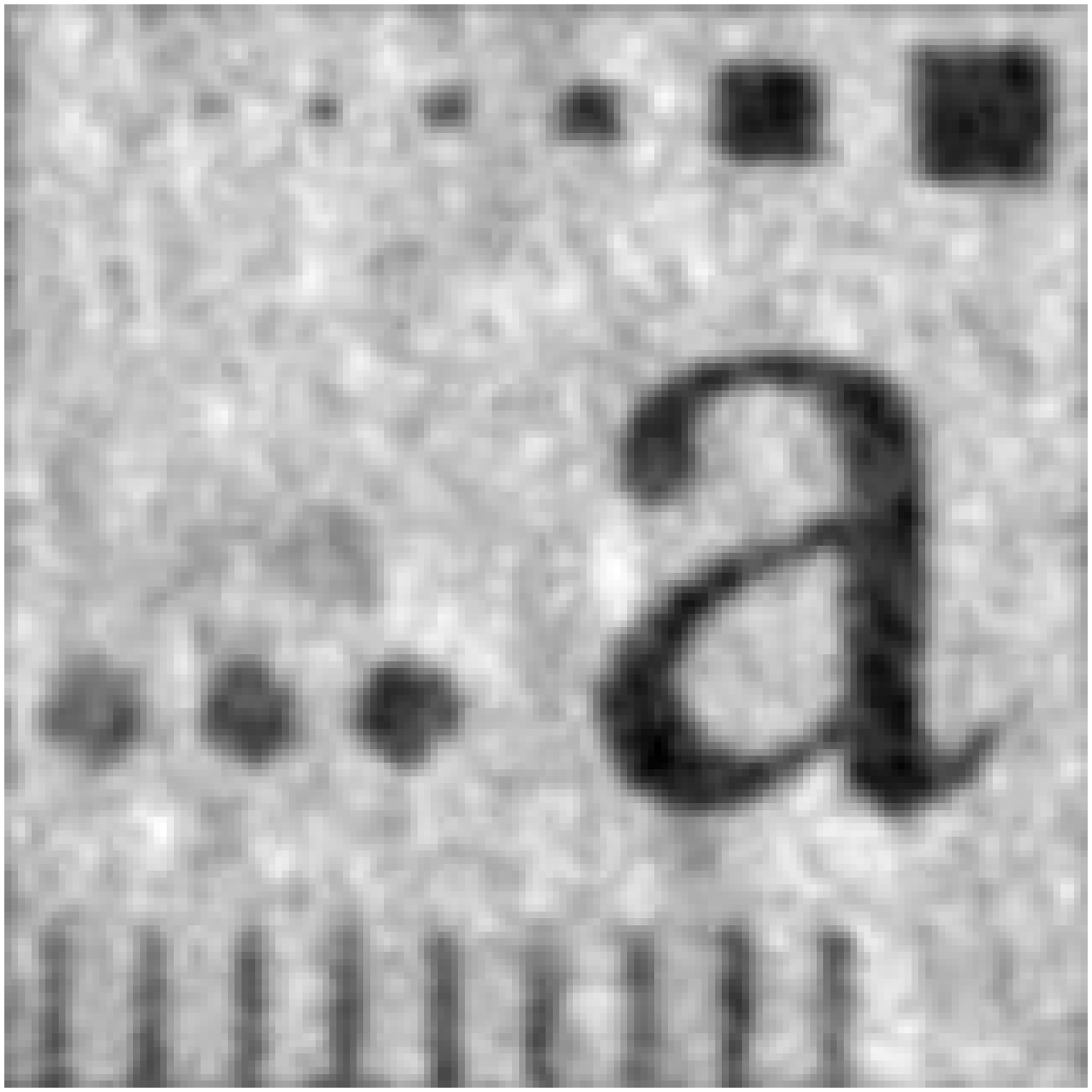}\label{f2j}}
\subfigure[]{
\includegraphics[width=0.1\textwidth]{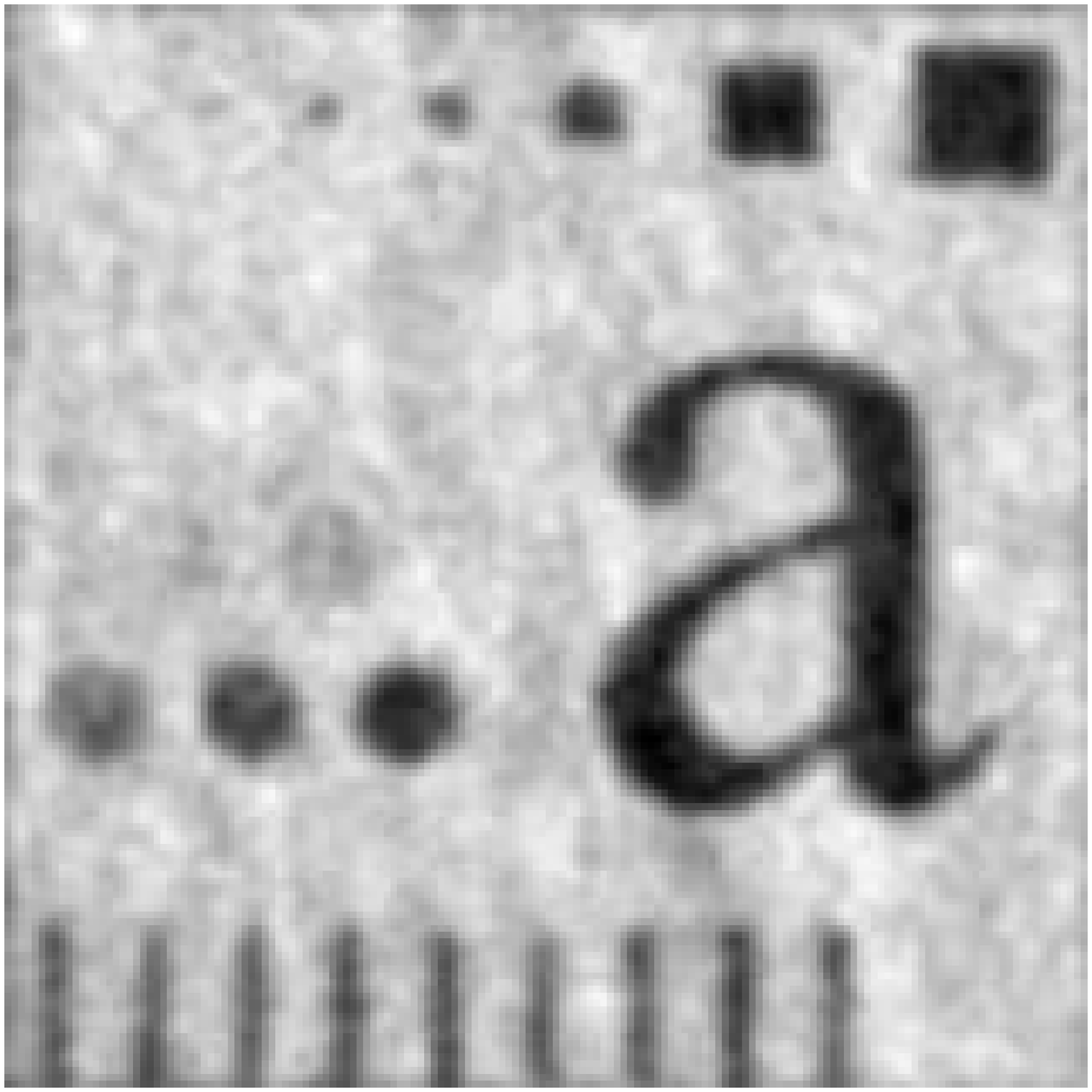}\label{f2k}}
\subfigure[]{
\includegraphics[width=0.1\textwidth]{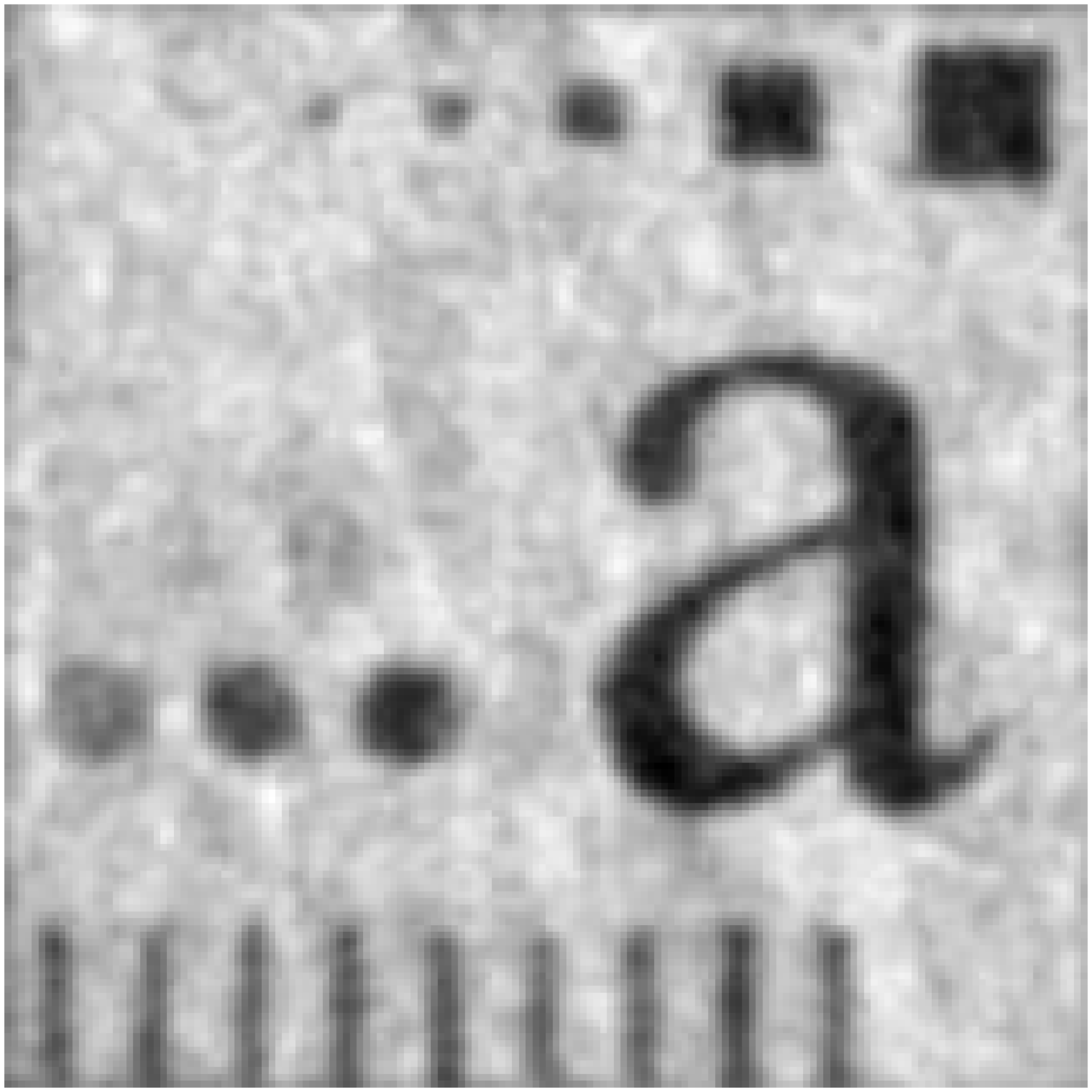}\label{f2l}}
\includegraphics[width=0.02759\textwidth]{figgray2.eps}\\
\caption{\label{f2} Top row: (a) Digital mask 1. (b) DGI and (c) GI images, (d)
CI positive minus negative image,
$\langle{I}_{R}\rangle_{+}-\langle{I}_{R}\rangle_{-}$, all from $140 000$
frames. Center row: (e) Positive TCDGI image $\langle I_{R}\rangle_{+}^{diff}$,
(f) negative TCDGI image $\langle I_{R}\rangle_{-}^{diff}$, (g) positive TCDGI
image minus reference signal average $\langle\delta {I}_{R}\rangle_{+}^{diff}$,
(h) negative TCDGI image minus reference signal average
$\langle\delta{I}_{R}\rangle_{-}^{diff}$, all from $69 600$ frames. Bottom row:
TCDGI images obtained by $\langle I_{R}\rangle_{k^{+}}^{diff}-\langle
I_{R}\rangle_{-k^{-}}^{diff}$ with different thresholds. Number of frames
averaged: (i) $k_{3}$, $6000\times 2$, (j) $k_{2}$, $21 000\times 2$, (k)
$k_{1}$, $31 500\times 2$, (l) $k_{0}=0$, $69 600\times 2$.}
\end{figure}

\begin{figure}[!hbt]
\begin{picture}(0.1,0.1)(115,95)
\subfigure[]{\includegraphics[width=0.1\textwidth]{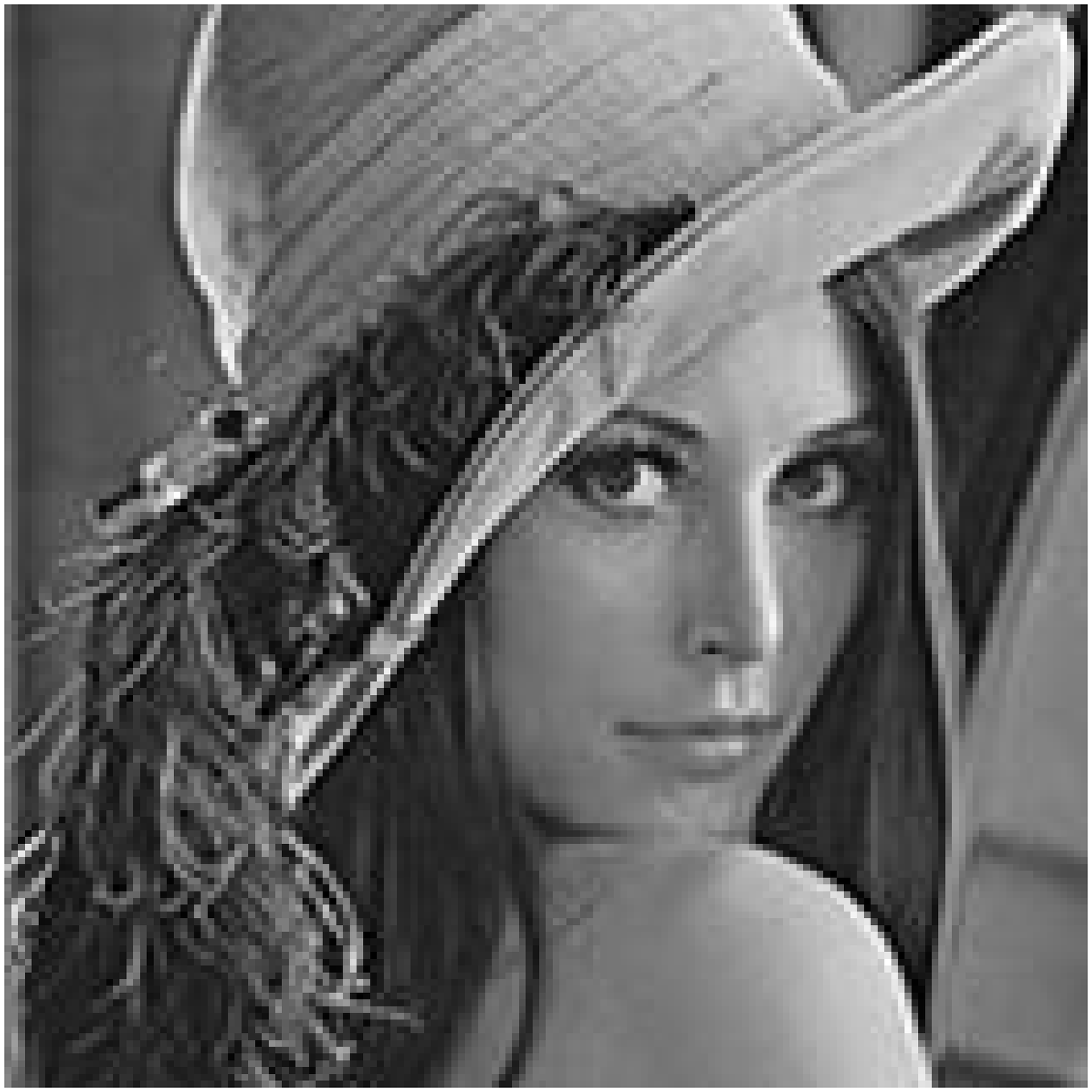}\label{f5a}}
\end{picture}
\flushright \subfigure[]{
\includegraphics[width=0.1\textwidth]{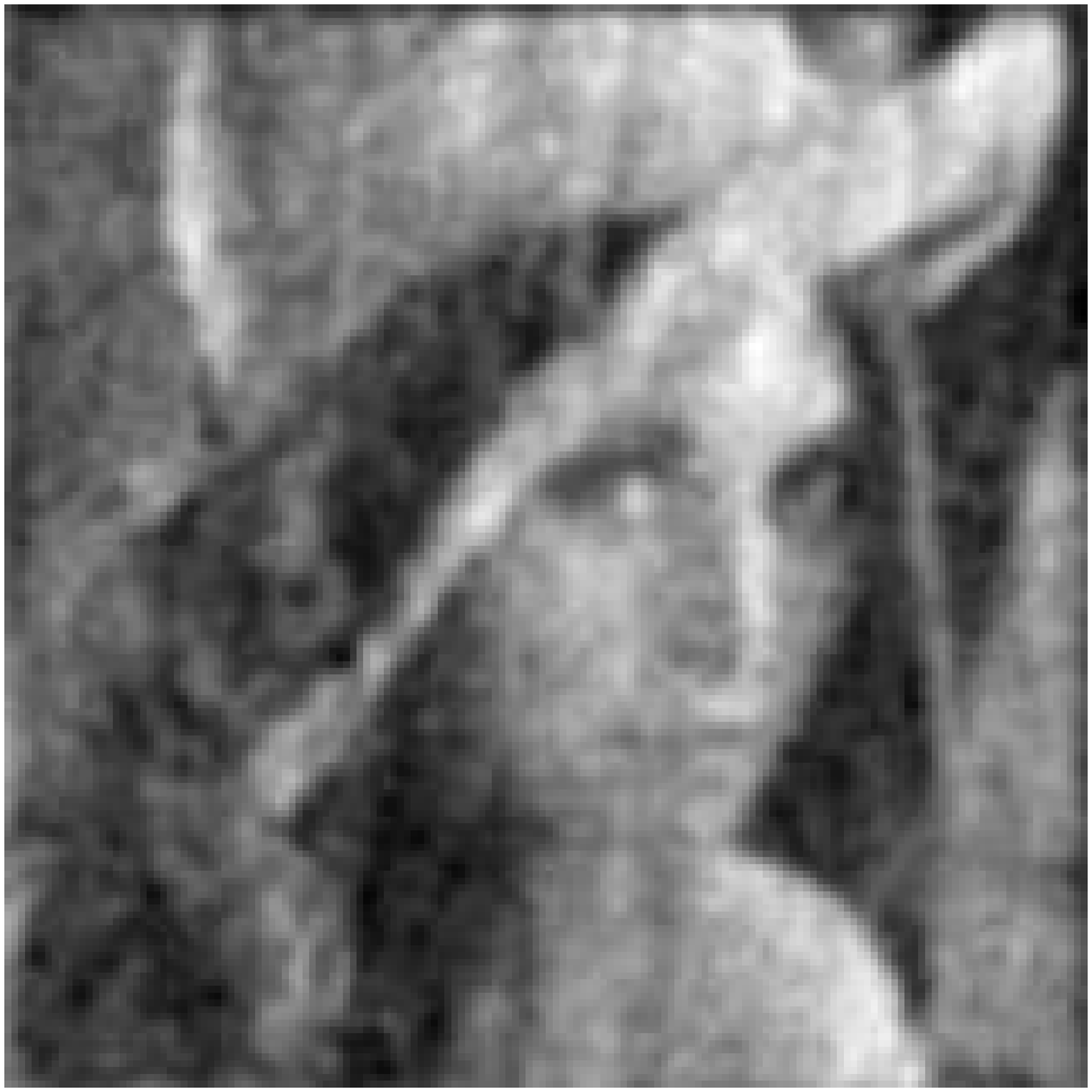}\label{f5b}}
\subfigure[]{
\includegraphics[width=0.1\textwidth]{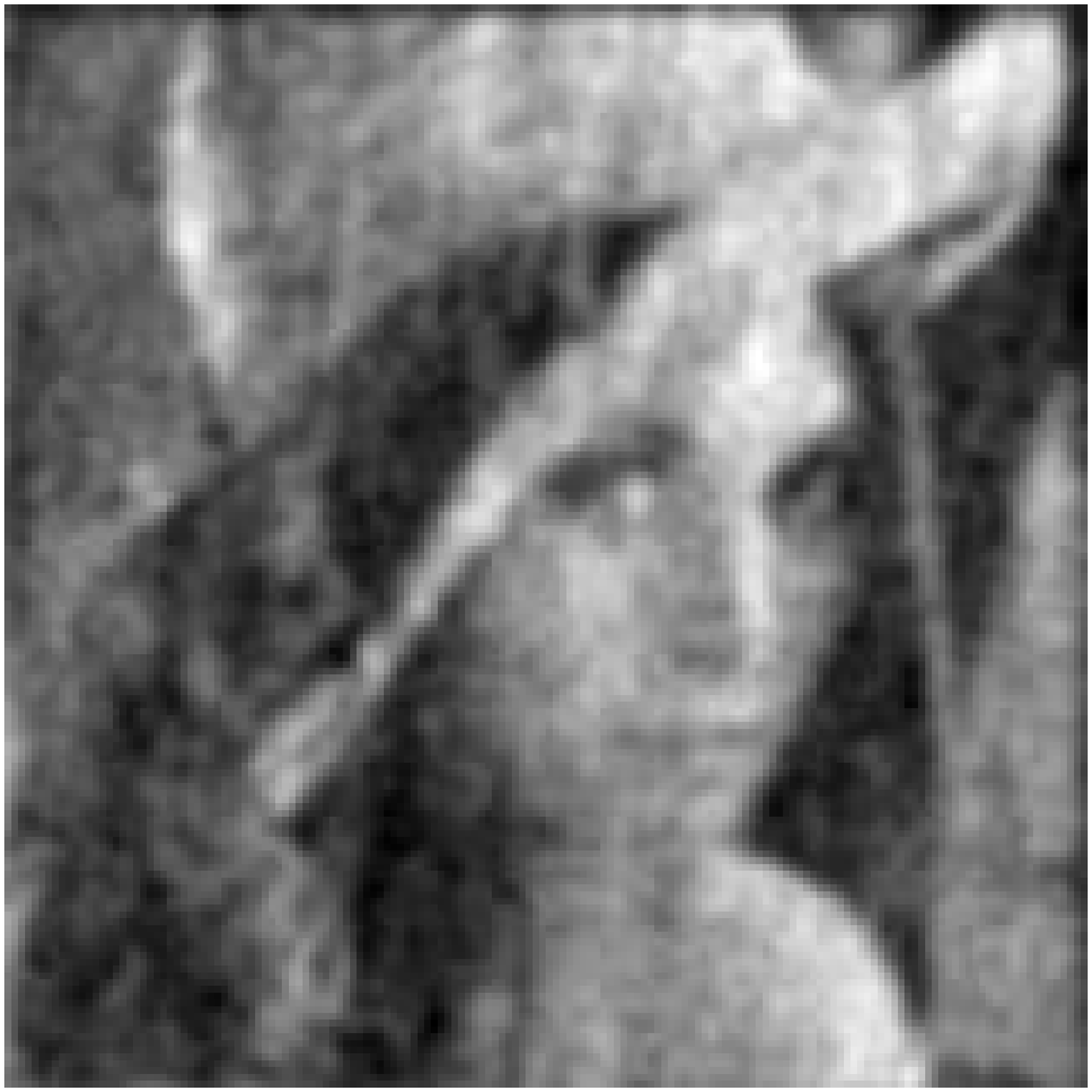}\label{f5c}}
\subfigure[]{
\includegraphics[width=0.1\textwidth]{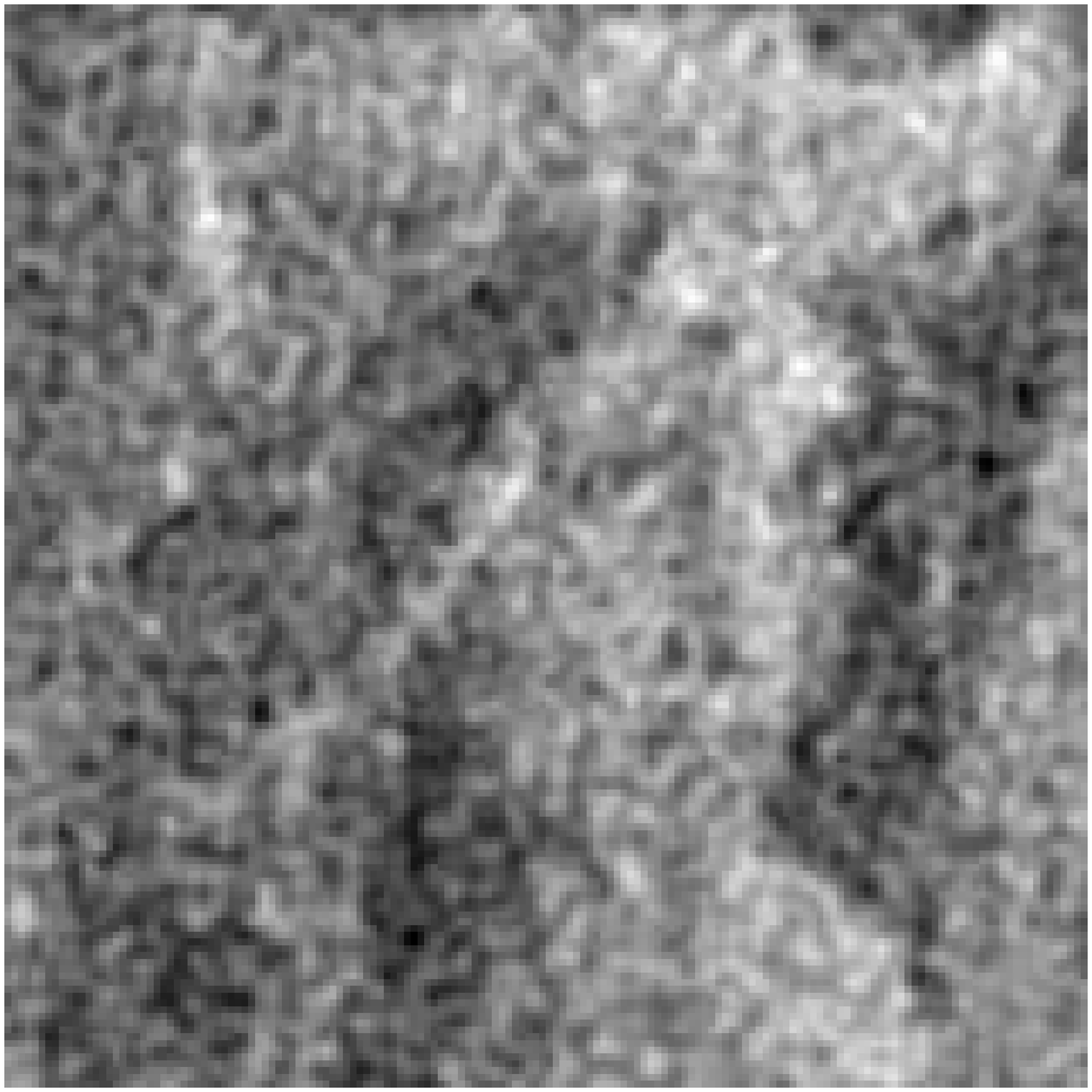}\label{f5d}}
\includegraphics[width=0.02759\textwidth]{figgray2.eps}\\
\subfigure[]{
\includegraphics[width=0.1\textwidth]{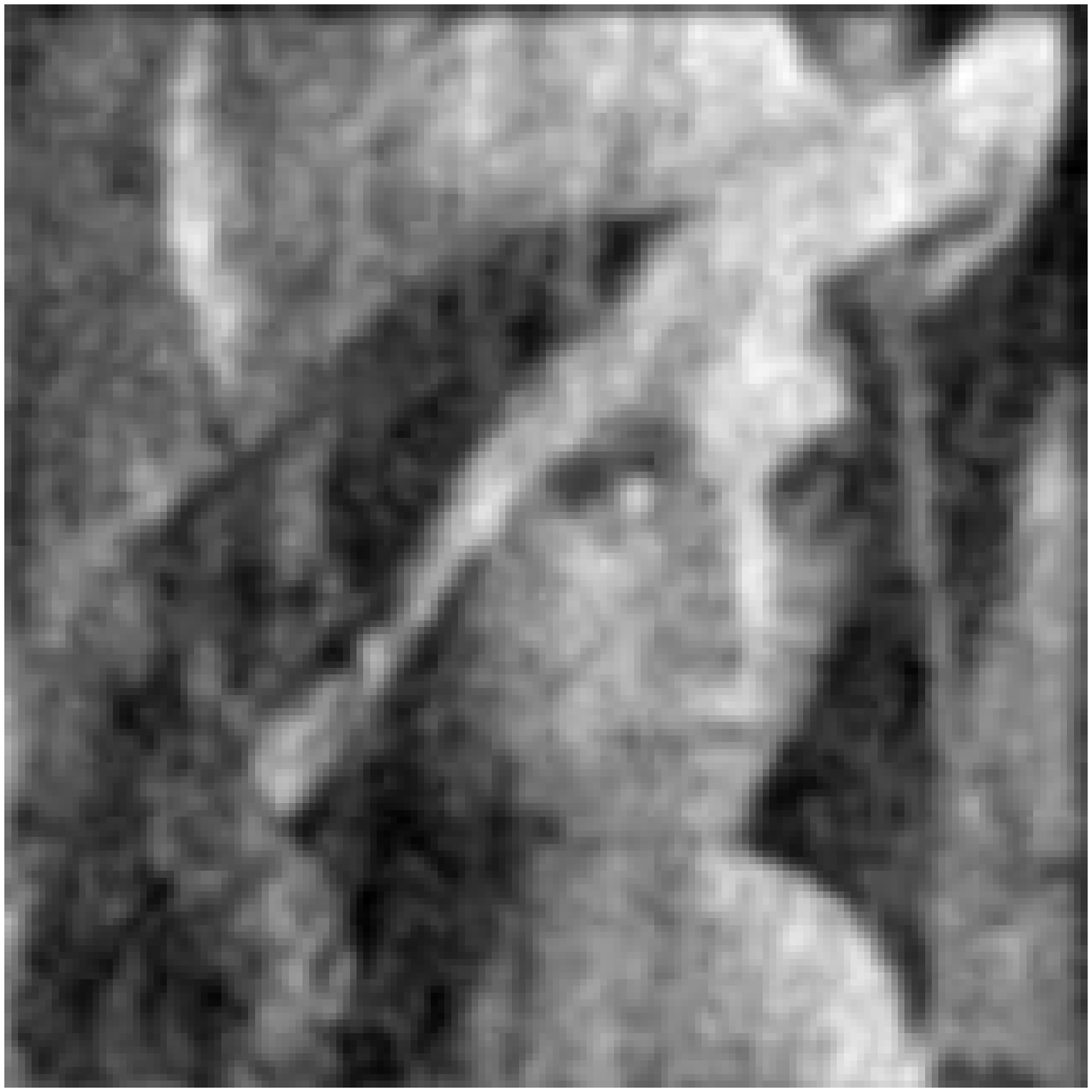}\label{f5e}}
\subfigure[]{
\includegraphics[width=0.1\textwidth]{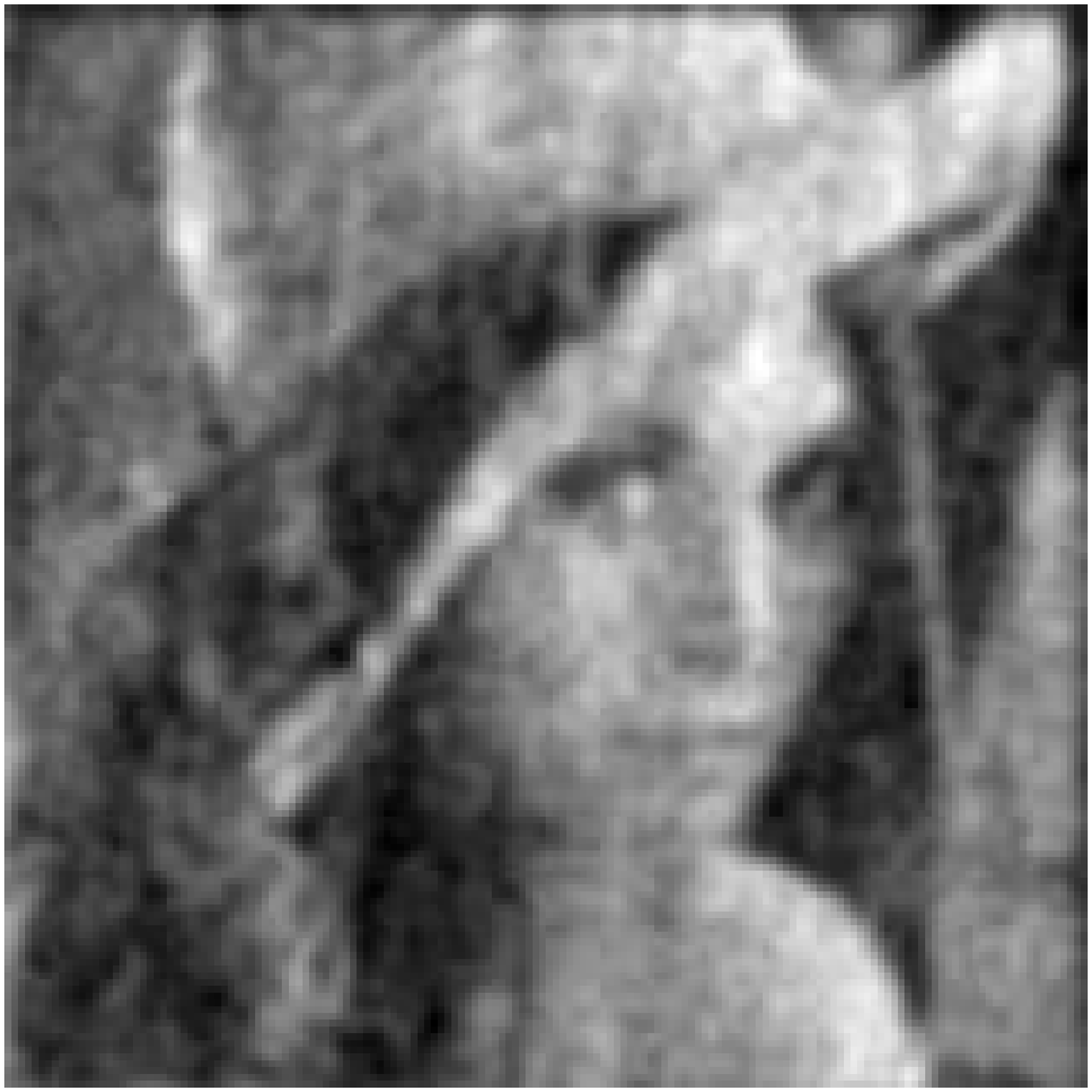}\label{f5f}}
\subfigure[]{
\includegraphics[width=0.1\textwidth]{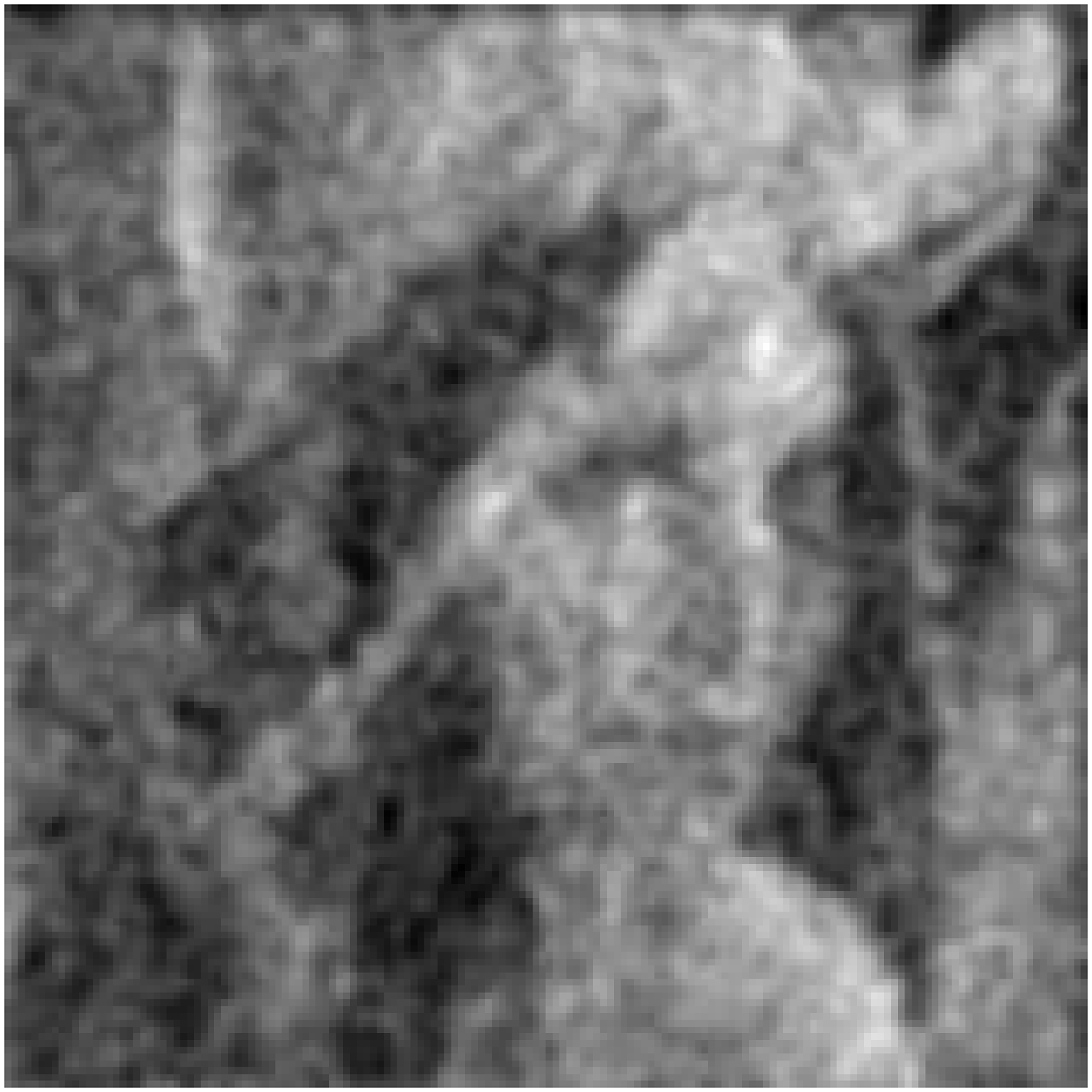}\label{f5g}}
\includegraphics[width=0.02759\textwidth]{figgray2.eps}\\
\caption{\label{f5}(a) Digital mask of Lena. Upper row: DGI images from (b)
$140 000$, (c) 71000, (d) 8000 frames. Lower row: TCDGI images from (e)
$69 000\times 2$, (f) $35 400\times 2$, (g) $3800\times 2$ frames for $k$ = 1, 2 and 3, respectively.}
\end{figure}

In the experiment, both cameras captured $1.4\times10^{5}$ frames with an
exposure time of $10^{-4}$ s. A square region in each detector
array, symmetric relative to the BS, of size $160\times160$ pixels was
selected, corresponding to the size of the object, which was a (virtual)
digital mask. Two different masks of the same size were used, one with strong
black-and-white contrast and the other with warm gray tones; their intensity
transmission functions $T_{1}(\bm{x})$ and $T_{2}(\bm{x})$ are shown in Figs.\
\ref{f2a} and\ \ref{f5a}, respectively. To obtain the bucket signal $S_{B}$, we
multiply the matrix of the mask with the corresponding intensity values
recorded by $\rm {D_B}$, pixel by pixel, and then sum over all the intensities
in the chosen region.  As in CI \cite{khl}, logical filtering is used to
divide the reference signals $I_{R}(\bm{x}_{R})$ into subsets that satisfy
specific conditions, and the image is then reconstructed by averaging over each
subset separately, but here for TCDGI, the condition depends on the fluctuation
of the differential signals.

Our theory adopts a classical formalism in which the shot noise is neglected,
and we simply set $I_{B}(\bm{x}_{B})=\alpha I_{R}(\bm{x}_{R})$ where the factor
$\alpha$ exists because of the imbalance of the beamsplitter and detectors.

In conventional GI, the image in terms of $T(\bm{x}_{R})$ is obtained by the
second-order correlation function of the intensity fluctuations of two
detectors \cite{t6,apl}
\begin{eqnarray}
\delta G^{(2)}_{GI}(\bm{x}_{R})=\langle\delta S_{B}\delta
I_{R}(\bm{x}_{R})\rangle\simeq C_{0}T(\bm{x}_{R}),
\end{eqnarray}
where $C_{0}=A_{coh}\langle I_{R}(\bm{x}_{R})\rangle\langle
I_{B}(\bm{x}_{B})\rangle$ is a constant, $T(\bm{x}_{B})$ denotes the intensity
transmission function of the object, and the bucket signal of the object arm is
defined as $S_{B}=\int I_{B}(\bm{x}_{B})T(\bm{x}_{B})d^{2}\bm{x}_{B}$, $\delta S_{B}=S_{B}-\langle S_{B}\rangle$, and $\delta I_{R}(\bm{x}_{R})=I_{R}(\bm{x}_{R})-\langle I_{R}(\bm{x}_{R})\rangle$.

To implement DGI \cite{DGI}, which can dramatically enhance the SNR, we need to
define the integrated reference detector signal $S_{R}=\int
I_{R}(\bm{x}_{R})d^{2}\bm{x}_{R}$. The differential bucket signal can be
written in an operative form as $S_{\Delta}=S_{B}-\frac{\langle
S_{B}\rangle}{\langle S_{R}\rangle}S_{R}$, and the quantity measured is
\begin{eqnarray}
\delta G^{(2)}_{16}(\bm{x}_{R})=\langle\delta S_{\Delta}\delta
I_{R}(\bm{x}_{R})\rangle\simeq C_{0}\delta T(\bm{x}_{R}), \label{diff}
\end{eqnarray}
in which the fluctuating part is $\delta
T(\bm{x}_{R})=T(\bm{x}_{R})-\overline{T}$, and $\overline{T}$ can be expressed
as $\langle S_{B}\rangle/\alpha\langle S_{R}\rangle$.

In the CI experiment \cite{khl}, all the reference frames are divided into two
subsets according to the sign of $\delta S_{B}$:
\begin{eqnarray}
\{I_{R}(\bm{x}_{R})|\delta S_{B}>0\},\ \{I_{R}(\bm{x}_{R})|\delta S_{B}<0\}.
\label{CI}
\end{eqnarray}
From each of the subsets, the image is then reconstructed merely by simple
averaging. We can obtain both positive and negative images from calculation of
the conditional averages $\langle I_{R}(\bm{x}_{R})\rangle_{+}$ and $\langle
I_{R}(\bm{x}_{R})\rangle_{-}$, respectively, which can be written as
\begin{eqnarray}
\langle I_{R}(\bm{x}_{R})\rangle_{\pm}
&&\equiv\langle I_{R}(\bm{x}_{R})\rangle+\langle \delta I_{R}(\bm{x}_{R})\rangle_{\pm}\nonumber\\
&&\simeq\langle I_{R}(\bm{x}_{R})\rangle+\langle \delta I_{R}(\bm{x}_{R})(1\pm\delta S_{B}/|\delta S_{B}|)\rangle\nonumber\\
&&\simeq\langle I_{R}(\bm{x}_{R})\rangle\pm{\langle\delta I_{R}(\bm{x}_{R})\delta S_{B}\rangle}/{\langle|\delta S_{B}|\rangle}\nonumber\\
&&\simeq\langle I_{R}(\bm{x}_{R})\rangle\pm{C_{0}T(\bm{x}_{R})}/{\langle|\delta
S_{B}|\rangle}, \label{pn}
\end{eqnarray}
Here we have assumed that the positive and negative frames are approximately
equal in number, and $\langle A/B\rangle\simeq\langle A\rangle/\langle
B\rangle$, as in Ref.\ \cite{optexp}.

We now divide the reference CCD signals into two subsets according to the sign
of $\delta S_{\Delta}=S_{\Delta}-\langle S_{\Delta}\rangle=S_{\Delta}$,
\begin{eqnarray}
\{I_{R}(\bm{x}_{R})|\delta S_{\Delta}>0\},\ \{I_{R}(\bm{x}_{R})|\delta
S_{\Delta}<0\}. \label{s-}
\end{eqnarray}
From Eqs.\ (\ref{diff}) and (\ref{pn}), we can thus obtain
\begin{eqnarray}
\langle I_{R}(\bm{x}_{R})\rangle^{diff}_{\pm}\simeq\langle
I_{R}(\bm{x}_{R})\rangle\pm{C_{0}\delta T(\bm{x}_{R})}/{\langle|\delta
S_{\Delta}|\rangle}, \label{dpn}
\end{eqnarray}
where $\langle\cdots\rangle_{+}^{diff}$ and $\langle\cdots\rangle^{diff}_{-}$
denote the averages of the positive and negative subsets, and correspond to the
positive and negative images of TCDGI determined by the sign of $\delta
S_{\Delta}$. In addition, it can be seen from Eqs.\ (\ref{pn}) and (\ref{dpn})
that if we normalize the average of the conditional reference intensity after
deducting the average of the reference signals, the image that we reconstruct
by CI and TCDGI is almost the same as that in conventional GI and DGI. However,
an even better way to reconstruct the image is to average all the normalized
information from the reference detector but with the negative signals inverted:
\begin{eqnarray}
\langle I_{R}(\bm{x}_{R})\rangle_{+}^{diff}-\langle
I_{R}(\bm{x}_{R})\rangle_{-}^{diff}=\frac{2C_{0}\delta
T(\bm{x}_{R})}{\langle|\delta S_{\Delta}|\rangle}, \label{p-n}
\end{eqnarray}
where the image is retrieved by only averaging the reference data. Compared
with DGI, the multiplication process is replaced by a logical filter process followed by simple addition,
thus computing time is saved. Furthermore, TCDGI has a higher SNR than that of
straightforward CI, which means that, with the same amount of data, TCDGI gives
a better image.

Next we discuss a more general scheme in which the selection condition is
modified to:
\begin{eqnarray}
\{I_{R}(\bm{x}_{R})|\delta S_{\Delta}>k\},\ \{I_{R}(\bm{x}_{R})|\delta
S_{\Delta}<-k\}, \label{tc}
\end{eqnarray}
where $k$ is an intensity threshold satisfying $0\leq k\leq\max\{|\delta
S_{\Delta}|\}$. The averages of the reference signals that satisfy the
threshold conditions are
$\langle I_{R}(\bm{x}_{R})\rangle^{diff}_{\pm
k^{\pm}}=\frac{1}{\beta}\left\langle I_{R}(\bm{x}_{R})\left(1\pm\frac{\delta
S_{B}\mp k}{|\delta S_{B}\mp k|}\right)\right\rangle$,
where $\beta=N_{k^{+}}/N_{0^{+}}$, with $N_{0^{+}}$ and $N_{k^{+}}$ being,
respectively, the number of frames satisfying $\delta S_{\Delta}>0$ and $\delta
S_{\Delta}>k$, respectively. The expressions $\langle\cdots\rangle_{+k^{+}}^{diff}$ and
$\langle\cdots\rangle_{-k^{-}}^{diff}$ denote the averages of the subsets
selected by the above two threshold conditions, and ${\pm k^{\pm}}$ means
plus-or-minus the real number greater than $+k$ (or less than $-k$). Thus, a
positive image can be obtained by subtracting these two averages:
\begin{eqnarray}
\langle I_{R}(\bm{x}_{R})\rangle^{diff}_{k^{+}}-\langle
I_{R}(\bm{x}_{R})\rangle^{diff}_{-k^{-}} \simeq C_{0}\delta T'(\bm{x}_{R}),
\label{tpn}
\end{eqnarray}
where $\delta T'(\bm{x}_{R})=\delta T(\bm{x}_{R})/\beta\langle|\delta
S_{\Delta}-k|\rangle$. In correlation imaging the main time consuming operation is the processing of all the big matrices of the reference frames, which is unavoidable even in DGI. In our TCDGI method, computation time is saved not just because we only use part of the matrices, but especially because we only need to add these matrices and then perform one minus operation, rather than having to multiply all of the differential bucket intensity signals one by one with the matching reference CCD matrix then take the average, as in the DGI protocol. Although the same original number of exposures frames must be taken and sorted, the sorting process can be regarded as almost instantaneous. For complex gray-scale objects, which we shall study below, much more imaging data is required, thus the advantage of less processing and computing time is even more pertinent.

We now present the experimental demonstrations of the above
methods. In the first experiment, we used a digital mask which is a widely used standard
in imaging processing [see Fig.\ \ref{f2a}]. To compare the quality of the
images obtained by different methods with the same standard, the gray scale of
every image is normalized within the interval $[0,1]$. This is achieved by
subtracting the minimum element value from each matrix element of the image
and then dividing the new matrix by the maximum element value. Figures \ \ref{f2b}
and \ref{f2c} show the images retrieved after averaging over $140000$ frames
by DGI and GI, respectively (gray bars are provided for comparison on the right
side of each row). In Fig.\ \ref{f2d}, we divide the reference frames into
positive and negative subsets in accordance with the sign of $\delta S_{B}$ in
Eq.\ (\ref{CI}) as in CI and then obtain the image by subtracting the negative
from the positive average,
$\langle{I}_{R}\rangle_{+}-\langle{I}_{R}\rangle_{-}$. Positive and negative
images obtained by TCDGI, where instead of using $\delta S_{B}$, the sign of
$\delta S_{\Delta}$ is used to divide the reference signals, are shown in
Figs.\ \ref{f2e} and \ref{f2f}, respectively; here 69600 frames were taken.
Both positive and negative images are much better than in GI, although there
are still blurred striations due to the rotation of the ground glass plate,
which reflect the nonconstant background $\langle I_{R}(\bm{x}_{R})\rangle$ in
Eq.\ (\ref{pn}). If the number of frames is large enough, this term will tend
to a constant, and the quality will improve. However, we can eliminate the
background noise without increasing the number of measurements by directly
subtracting the total average of the reference signals, to obtain
$\langle\delta {I}_{R}\rangle_{+}^{diff}$ and
$\langle\delta{I}_{R}\rangle_{-}^{diff}$. As we can see from Figs.\ \ref{f2g}
and \ref{f2h}, the background has been almost completely eliminated, and both
positive and negative images are much better and clearer than in all previous
methods.

From Eq.\ (\ref{tpn}) we can see that, in TCDGI, even if only part of the
reference signals are chosen according to various threshold values, we can
still retrieve relatively high quality images. Note that there is a one-to-one
correspondence between the intensity threshold $k$ and the number of selected
frames; the latter decreases as $k$ increases. The TCDGI images in the bottom
row of Fig.\ \ref{f2} are reconstructed from different threshold values and
frame numbers, again obtained by subtracting the negative from the positive
image, see Eq.\ (\ref{tpn}). Figures\ \ref{f2i} to \ref{f2l} are images
retrieved for $k_{3}>k_{2}>k_{1}>k_{0}=0$ from 6000, 21 000, 31 500 and 69 600 frames, respectively, from each $\delta S_{\Delta}>k_{i}$
and $\delta S_{\Delta}<-k_{i}$ ($i=0, 1, 2, 3$) section. Although Fig.\ \ref{f2i} does not appear to
have such good contrast as in Fig.\ \ref{f2l}, it was reconstructed from only
6000$\times 2$ frames as compared with 69 600$\times 2$. On the other hand, the
first smallest black square in the top left hand corner of the mask is quite
visible in\ \ref{f2i} but almost indiscernible in\ \ref{f2l}. In all cases,
subtracting the negative frames always gives a greatly improved image, with
much of the background noise removed.

\begin{figure}[!hbt]
\centering {\includegraphics[width=0.35\textwidth]{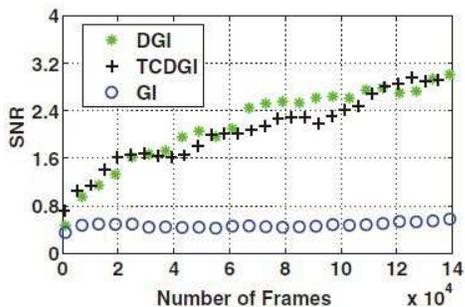}}
\caption{\label{f3}(Color online) SNR vs. number of reference frames.}
\end{figure}

To provide a quantitative comparison of the image quality obtained by various
methods, we define the SNR \cite{snr1, snr2} as:
\begin{eqnarray}
{\rm {SNR}=\frac{Signal}{Noise}}=
\frac{\sum_{i,j=1}^{M,N}[T_{0}(i,j)-\overline{T}_{0}]^2}{\sum_{i,j=1}^{M,N}[T(i,j)-T_{0}(i,j)]^2},
\label{SNR}
\end{eqnarray}
where $T_{0}(i,j)$ and $T(i,j)$ are the transmission matrices of the object
mask of size $M \times N$ and the retrieved image, respectively, and
$\overline{T}_{0}=(MN)^{-1}\sum_{i,j=1}^{M,N}T_{0}(i,j)$. Generally speaking,
we would expect the SNR to improve with the number of reference frames averaged
over. The SNRs corresponding to $k_{3}$, $k_{2}$ and $k_{1}$ in Figs.\
\ref{f2i}, \ref{f2j} and \ref{f2k} are 2.12, 3.21 and 3.24, respectively. But
for a fair comparison, we should calculate the SNR for the same $k$ value. In
Fig.\ \ref{f3} we plot the SNR versus number of exposures for a threshold value
of $k=0$. We see that the SNR of DGI (stars) and TCDGI (crosses) are almost
the same, and both are better than that of GI (circles), which is in good
agreement with the results predicted by our theory. The highest points measured
by DGI, GI and TCDGI correspond to the images in Figs.\ \ref{f2b}, \ref{f2c} and
\ref{f2l}, respectively.

\begin{figure}[!hbt]
\centering {\includegraphics[width=0.34\textwidth]{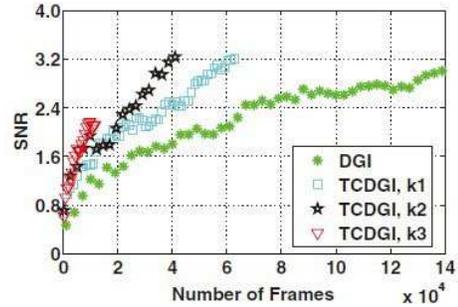}}
\caption{\label{f4}(Color online) SNRs of DGI and TCDGI of different thresholds
against the number of frames.}
\end{figure}

To demonstrate the advantages of TCDGI, the SNRs for different thresholds
against the number of frames are given in Fig.\ \ref{f4}. It is evident that as
the threshold $k$ increases, the SNR increases faster, but is limited by the
number $N_{k^{+}}$ of reference frames available for averaging. As the
computational time is approximately proportional to the number of frames, our
TCDGI method is able to retrieve an image of better quality than DGI but with
much less data manipulation and computing time. It is interesting that the SNR
curves all exhibit some slight oscillations, as also observed in Ref.
\cite{Clemente-SPGH}. To our knowledge, this is partly due to experimental
errors, and partly because the information in the CCD matrix frames is actually
redundant for retrieving the object.

So far there are few GI experiments on objects with continuously changing gray
tones, due to the immense number of exposures required. The mask that we used
in the second experiment is the well-known photo of Lena that is widely used in
traditional image processing tests. The images obtained by different methods
are shown in Fig.\ \ref{f5}, and their corresponding SNRs in the Table. We can
see clearly that, for the same number of frames, the images retrieved by DGI
(upper row) are inferior in quality to those retrieved by TCDGI (lower row).
Moreover, with only $7600$ exposures, we can still distinguish the TCDGI image
in Fig.\ \ref{f5g}, while the DGI image from $8000$ frames (Fig.\ \ref{f5d}),
is almost drowned in the noise. It is interesting that the maximum SNR value of
2.49, obtained by TCDGI, (1.95 for DGI) is not given by the highest number of
exposures ($69 000\times 2$ in Fig.\ \ref{f5e}, which has an SNR of 2.43) but by
$35 400\times 2$ exposures (Fig.\ \ref{f5f}); the reason for this is, however,
unclear at present, and deserves further analysis.

\begin{table}
\caption{\label{tab} SNR of the images in Fig.\ \ref{f5}.}
\begin{ruledtabular}
\begin{tabular}{c |c c c |c c c }
 Mask &  & DGI & & & TCDGI \\
\hline
\ref{f5a} &  \ref{f5b} &   \ref{f5c}&   \ref{f5d} & \ref{f5e}  & \ref{f5f}  & \ref{f5g}\\
\colrule
$\infty$\footnote{SNR of the mask is infinite, in accordance with Eq.\ (\ref{SNR}).} & 2.10  & 1.95  & 0.75   & 2.43 & 2.49  & 1.65 \\
\end{tabular}
\end{ruledtabular}
\end{table}

In conclusion, we have presented a new imaging technique called TCDGI
by which we can retrieve the image of an object through conditional averaging
of the spatial intensity together with inversion of the negative signals, using
only the reference detector data. This method can dramatically enhance the
SNR compared with conventional GI and CI, especially for objects with rich gray
tones. Moreover, under the same conditions but with appropriate choice of
threshold values, the retrieved image can be better than DGI, whilst being simpler to process, and requiring
less data manipulation and computing time. It is shown that the major contributions to the retrieved
image come from the exposures with the largest intensity fluctuations. We
believe that this new technique, which combines the advantages of CI and DGI,
may become a standardized method in real applications where conventional
imaging and GI protocols do not work well.

This work was supported by the National Basic Research Program of China (Grant
No.2010CB922904), the National Natural Science Foundation of China (Grant
No.60978002) and the Hi-Tech Research and Development Program of China (Grant
No.2011AA120102). The authors are grateful for valuable discussions with
Xue-Feng Liu and Xu-Ri Yao.

\end{document}